\documentclass[12pt,preprint]{aastex}

\newcommand{\ssec}{$^{\rm s}$}                         
\newcommand{\smin}{$^{\rm m}$}                         
\newcommand{\shour}{$^{\rm h}$}                        
\newcommand{\dsec}{\hbox{$.\!\!^{\rm s}$}}             
\newcommand{\dasec}{\hbox{$.\!\!^{\prime\prime}$}}     
\newcommand{\damin}{\hbox{$.\!\!^{\prime}$}}           
\newcommand{\sdeg}{$^{\circ}$}                         
\newcommand{\ddeg}{\hbox{$.\!\!^\circ$}}               
\newcommand{\nh}{$N_{\rm{H}}$ }                         
\newcommand{\av}{$A_{V}$ }                         
\newcommand{\ak}{$A_{K}$ }                         

\shorttitle{GC Star Clusters}
\shortauthors{Law \& Yusef-Zadeh}

\begin{document}


\title{X-Ray Observations of Stellar Clusters Near the Galactic Center}


\author{C. Law and F. Yusef-Zadeh}
\affil{Department of Physics and Astronomy, Northwestern University, Evanston, IL 60208}


\begin{abstract}
We report the first detection of X-ray emission from the Quintuplet star cluster and compare its X-ray emission to that of the Arches star cluster.  Four point sources are significantly detected near the core of the Quintuplet cluster with a total, absorption-corrected luminosity of $\sim1\times10^{33}$ ergs s$^{-1}$.  Diffuse, thermal emission is also detected near the core of the Quintuplet cluster with an absorption-corrected luminosity of $\sim1\times10^{34}$ ergs s$^{-1}$.  We analyze the diffuse and point-like emission from the Arches and Quintuplet and discuss the possibility that they are host to cluster wind outflows.  We also present the results of our search for X-ray emission from candidate star clusters in the Galactic center (GC) region.  We use extinction estimated by near-IR colors and X-ray spectral fits, as well as other IR properties, to determine if the candidate clusters are new, GC star clusters.  We find that three of the six candidate clusters found toward the GC are likely foreground clusters, two of the candidate clusters are not detected in the X-ray data, but have near-IR extinctions consistent with a GC location, and one of the candidate clusters has X-ray and near-IR extinctions consistent with being in the GC.  The X-ray and IR emission from the candidate clusters is compared to the known, massive, GC star clusters.
\end{abstract}




\section{Introduction}
The Galactic center (GC) region is host to some spectacular stellar clusters.  The Arches, Quintuplet, and central clusters are all young and among the densest in the galaxy.  The distribution of these clusters within 30 pc (12\damin5 assuming a distance of 8.5 kpc) of Sgr A* may not be coincidental.  The ambient conditions of the GC region are known to be extreme (e.g., high gas and stellar densities, intense ionizing flux;  Morris \& Serabyn 1996) and should have a significant effect on the process of star formation.

The currently known massive, GC star clusters (i.e. the Arches, Quintuplet, and IRS 16 clusters) were discovered in surveys in the near-IR (Becklin \& Neugebauer 1975; Nagata et al.\ 1990; Cotera et al.\ 1996), where interstellar extinction is much less than at visible wavelengths.  Although the near-IR is an excellent regime in which to search for extincted stellar emission, the GC is particularly challenging region to survey because of confusion with stars lying along the line of sight, as well as the nonuniform distribution of extinction found there.  The Arches cluster was recently detected at X-ray wavelengths (Yusef-Zadeh et al.\ 2002, hereafter Y02), where the emission is thought to originate from the shocked gas in the collision between individual stellar winds, as had been modeled specifically for the Arches (Cant\'{o} et al.\ 2000; Raga et al.\ 2001).  These studies suggest that cluster wind emission may provide a new window into the physics of dense star clusters, although it can be difficult to separate putative cluster wind emission from unresolved stellar cluster members.

The Quintuplet cluster is another of these unusual, massive, GC star clusters.  The Quintuplet, named after its brightest five stars (Nagata et al.\ 1990), is unusually dense and is host to several massive, windy, Wolf-Rayet stars.  However, the Quintuplet is somewhat less massive and dense than the Arches cluster, suggesting that it has been more dissolved by GC tidal forces (Kim et al.\ 1999; Portegeis-Zwart et al.\ 2001).

Recent theoretical and observational studies have suggested that there may be many more undiscovered massive star clusters in the GC region.  \emph{N}-body simulations of these dense, young clusters have suggested that star clusters such as the Arches dynamically evolve rapidly, dissolving into the stellar background within 10--20 Myr (Portegeis-Zwart et al.\ 2001;  Kim et al.\ 1999; Kim \& Morris 2003).  Thus it was suggested that there may be another 10--50 massive clusters distributed within the central 150 pc at various stages of dissolving into the background stellar field.  On the observational side, Dutra \& Bica (2000, 2001; hereafter DB00, DB01) and Dutra et al.\ (2003) searched Two-Micron All Sky Survey (2MASS) images for Arches-like objects to make a list of candidate GC clusters.  Their results were consistent with rough estimates of the number of clusters predicted by Portegeis-Zwart et al.\ (2001).

This paper discusses the use of X-ray and IR properties of the Arches and Quintuplet clusters to examine whether the cluster candidates identified by DB00 \& DB01 are in the Galactic center region.  In \S \ref{known}, we extend our previous analysis of the emission from the Arches cluster and present the first detection of X-ray emission from the Quintuplet cluster.  Section \ref{candidate} presents our analysis of the candidate star clusters found by Dutra \& Bica;  in particular, X-ray and near-IR extinction are used to determine whether the candidate clusters are in the Galactic center.  We conclude by comparing the candidate clusters to the known GC clusters and commenting on the population of massive star clusters in the GC.

\section{Observations and Data Reduction}
\label{obs}

\begin{center}\emph{a) IR}\end{center}

Near-IR photometry ($J$, $H$, and $K_s$ bands) is taken from the 2MASS All-Sky Point Source catalog\footnote{Available at http://www.ipac.caltech.edu/2mass/releases/docs.html.} and near-IR images from the 2MASS image server.  Mid-IR images (8-21 $\mu\rm{m}$) are from the \emph{Midcourse Space Experiment} (\emph{MSX}) (Egan \& Price 1996).  Quality of near-IR photometry was measured by the 2MASS ``ph\_qual'' and ``rd\_flg'' flags (see previous footnote). Near-IR photometry (for color-color diagrams) was used if the ``ph\_qual'' flag was A, B, C, or D;  a source was considered to have some flux measured if the source had ``rd\_flg'' equal to 1, 2, 3, 4, 6, or 9.

The near- and mid-IR properties of the known and candidate clusters are summarized in Table \ref{properties}.  Aside from generic properties (position, size), diagnostics of intrinsic cluster properties were created for cluster comparison.  Column (5) of Table \ref{properties} gives a rough measure of the projected stellar density of the cluster, relative to the background.  This was found by comparing the number of 2MASS sources with any flux measured in the $K_s$ band in the cluster core, divided by the number of sources in an annulus from 1--3 times the cluster core.  The cluster ``core'' region was defined by visually searching the $K_s$ images for the highest concentration of sources near the locations given by DB00 and DB01.  Column 6 shows the number of sources with $H-K_s$ color consistent or inconsistent with GC values.  Extinction toward the inner degree is typically \ak$>2$ (Dutra et al.\ 2003), which is equivalent to $E_{H-K_{s}}>1.1$ (Cardelli et al.\ 1989).  At this extinction, 2MASS should be able to detect all GC supergiants (and possibly some giants) in the $H$ and $K_s$ bands, giving us a range of intrinsic $H-K_{s}$ colors from 0--0.3 (Cotera et al.\ 2000) and reddened values of $>$1.1--1.4.  Assuming these sources are supergiants and giants, sources with $H-K_{s}>1.4$ are probably at or beyond the GC, while sources with $H-K_{s}<1.1$ are probably in the foreground.

\begin{center}\emph{b) X-Ray}\end{center}

All X-ray observations were conducted by the \emph{Chandra X-Ray Observatory} (Weisskopf et al.\ 2000) with the Advanced CCD Imaging Spectrometer (ACIS) in imaging mode.  Two 50 ks observations were obtained from the \emph{Chandra} archive; ObsID 242 observed the Sgr A region (Baganoff et al.\ 2003) and ObsID 945 was centered near $l\sim0\ddeg2$ (Yusef-Zadeh et al.\ 2002).  In addition, the \emph{Chandra} GC Survey, as described in Wang, Gotthelf, \& Lang (2002), observed a $2$\sdeg$\times0\ddeg8$ region centered near Sgr A.  Each observation had a duration of $\sim$11 ks;  overlapping observations gave an exposure of about 22 ks for most of the GC region.  

All observations were reduced in an identical manner, using CIAO 2.2.1 \footnote{See http://cxc.harvard.edu/ciao.} and CALDB 2.12.  Charge transfer inefficiency corrections were applied, as described in Townsley et al.\ (2000).  Periods of high background flux were filtered by studying the light curve of blank portions of each observation;  only ObsID 242 and 945 were effected by background flares, reducing their usable exposure times by about 8 and 2 ks, respectively.  Finally, an error in the \emph{Chandra} data processing pipeline required a correction to the pointing solution.\footnote{See http://cxc.harvard.edu/cal/ASPECT/celmon/.}  This is the only astrometric correction applied for the work presented here, aside from the corrections discussed for ObsID 945 (which covers the Arches and Quintuplet clusters;  see Table \ref{properties}).  All spectral modeling performed without correcting for the reduction in low-energy quantum efficiency of ACIS.  After this analysis was completed, we compared spectral fits with and without the low-energy correction for a few of our spectra and found no significant difference in derived spectral parameters.

X-ray sources were detected with the \emph{wavdetect} tool of CIAO (Freeman et al.\ 2002).  The 3 $\sigma$ X-ray flux (i.e., detection) limits, in units of counts cm$^{-2}$ s$^{-1}$, were evaluated using equation (2) of Muno et al.\ (2003):

$S = \frac{n^2_\sigma}{2} \frac{1}{AT} \left[1+\left(1+\frac{8ba}{n^2_\sigma}\right)^{1/2}\right]$

\noindent where $n_\sigma$ is the signal to noise ratio of the detection limit, $b$ is the background count density, $a$ is the background area, $A$ is the effective area of the detector, and $T$ is the exposure time of the observation.  Since all regions studied here were observed multiple times, Table \ref{properties} shows the effective \emph{total} flux limit.  As shown above, the flux limit varies as the inverse of the exposure time, thus the total X-ray flux limit is equal to the inverse of the sum of the inverse flux limits of individual observations, that is:  $S_{\rm{tot}}^{-1}=\sum_{i} S_{i}^{-1}$.  Although CIAO's source detection algorithms can only be run on individual observations (as they are sensitive to the size and shape of the PSF), we find that these effective X-ray flux limits are consistent with the actual flux limits of the point source detection algorithms (compare Tables \ref{properties} and \ref{xirtab}).

Table \ref{xirtab} lists all X-ray sources detected near the known and candidate GC clusters, and any possible IR correlated sources.  In total, 14 sources are detected with greater than 2.5 $\sigma$ confidence within roughly two core radii of the candidates.  For this study we focus on the most likely X-ray/IR correlations, which are shown in boldface.  These X-ray sources (1) have more confident detections ($>$3 $\sigma$), (2) are located within the cluster core (i.e., distance$<$r$_{\rm{core}}$), and (3) have IR correlations within the nominal $1\arcsec$ positional error of \emph{Chandra} data.  For each X-ray source detected, Table \ref{xirtab} shows X-ray properties (flux, detection significance) and potential near-IR counterparts.  Officially, \emph{Chandra} source names preceed the J2000 coordinates with ``CXOGCJ'';  2MASS source names are preceeded by ``2MASS.''  X-ray flux is calculated from the \emph{wavdetect}-measured net counts, divided by the unnormalized exposure map for that position on the detector.  Count fluxes can be converted to energy fluxes (units of ergs cm$^{-2}$ s$^{-1}$) by assuming a spectral shape for the 0.5--8 keV band.  For the typical absorption and dust scattering toward the GC, a $\Gamma\sim1$ power law will have a conversion factor of $7.8\times10^{-9}$ ergs count$^{-1}$, while a thermal plasma model (MEKAL;  Liedahl et al.\ 1995) has conversion factors ranging from 4--8$\times10^{-9}$ ergs count$^{-1}$ for $kT=0.5$--3 keV.  \emph{Chandra's} absolute astrometric accuracy is about $1\arcsec$;  nominally, this is the largest offset expected between an X-ray source and an IR couterpart.  However, Feigelson et al.\ (2002) found that faint \emph{Chandra} sources observed $5\arcmin$ off-axis may have an additional $1\arcsec$ error in \emph{wavedetect} positions, because of distortions introduced by the PSF.

\section{X-ray Emission from Known Stellar Clusters}
\label{known}

\subsection{The Arches Cluster}
\label{arches}
\begin{figure}
\plotone{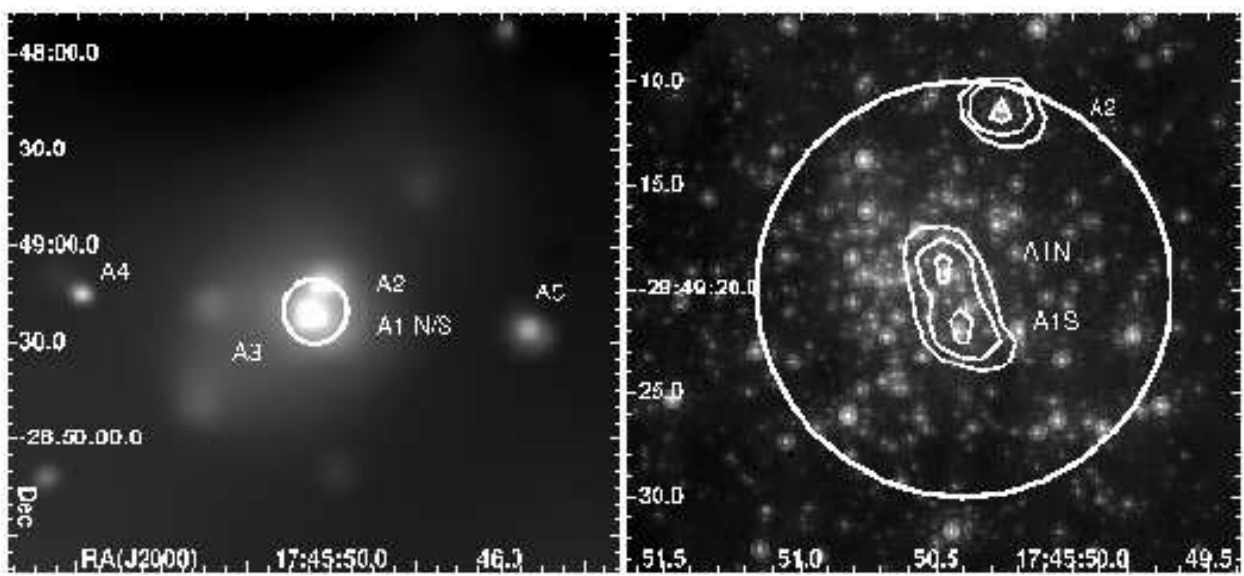}
\caption{(\emph{Left}):  Adaptively smoothed image of all \emph{Chandra} observations of the Arches star cluster from 0.5--8 keV.  Circles in both images show the extent of the cluster core, as listed in Table \ref{properties}. (\emph{Right}):  \emph{HST}/NICMOS image (Figer et al.\ 1999b) with X-ray contour levels of 0.7, 1.4, and 2.8 counts overlaid.  The X-ray contours were taken from the highest resolution image of the region (ObsID 2276), which resolved A1 into a north and south component. \label{archesfig}}
\end{figure}

The Arches cluster is one of the densest and most massive star clusters in the Milky Way and is located 30 pc in projection from Sgr A* (Cotera et al.\ 1996).  Figure \ref{archesfig} shows X-ray and near-IR images of the cluster.  The first observation of the cluster found the following sources:  A1 in the cluster core, A2 about $10\arcsec$ north of the core, the diffuse halo A3 emission, and two faint sources located $\sim1\arcmin$ east and west of the cluster core, A4 and A5, respectively (Y02).  Here we present the most recent high-resolution, X-ray observation of the Arches cluster, using the \emph{Chandra} GC Survey data.  The new images show that A1 is resolved into two components, A1N and A1S (coordinates are given in Table \ref{xirtab}).  After applying the corrections discussed in \S \ref{obs}, all three sources (A1N, A1S, and A2) are found to be roughly 0\dasec5 north of the nearest IR sources of Figer et al.\ (2002).  This suggests that the pointing for the \emph{Chandra} observation is in error;  a systematic offset of 0\dasec5 in the \emph{Chandra} pointing is consistent with the absolute astrometric accuracy of $\sim1\arcsec$.  The astrometry of the X-ray data is further corrected assuming these correlations were real;  this reduced the absolute astrometric uncertainty to 0\dasec6, relative to the positions in the Figer et al.\ (2002). \emph{HST} observation.

In addition to having IR counterparts, A1N and A1S are coincident with radio continuum sources AR4 and AR1 of Lang et al.\ (2001), respectively.  The radio emission is believed to form in the winds of massive O and Wolf-Rayet stars.  The correlation of the X-ray sources with previously identified stellar sources suggests the X-ray, IR, and radio emission have a similar origin, either in individual stellar winds or in colliding-wind binaries.  Overall, the Arches flux (within the $50\arcsec\times100\arcsec$ A3 boundary) is $\sim$40\% diffuse and $\sim$60\% point-like.  These fluxes do not account for the expected scattering of point-like emission by intervening dust.  Y02 suggested that A1 may be extended, as expected in the cluster wind models of Cant\'{o} et al\ (2000) and Raga et al.\ (2001); we instead find that the A1 source is resolved into two components.  As discussed in \S \ref{discussion}, wind collision emission may still be present in the cluster core, however the majority of the Arches X-ray flux is found in the A1N, A1S, and A2 sources, which are pointlike within the $\sim0\dasec5$ resolution of \emph{Chandra}.

The spectral analysis of the A1N, A1S, and A2 sources was undertaken in a number of ways, in order to separate intrinsic properties from effects of confusion and undersized spectral extraction regions.  The simplest approach was to fit each source with absorbed thermal models, while folding in a model fit to a background region.  This approach ignores the effect of the energy-dependent PSF, which reduces the number of high energy photons in the $\sim4\arcsec$ radius spectral extraction regions relative to the number of low energy photons.  With this small caveat, the fit values are: $kT_{\rm{A1N}}=1.6_{-0.3}^{+0.4}$ keV\footnote{All error ranges denote 90\% confidence intervals.}, $kT_{\rm{A1S}}=1.8_{-0.3}^{+0.4}$ keV, and $kT_{\rm{A2}}=2.2_{-0.5}^{+0.7}$ keV, with a dust-corrected\footnote{Dust along the line of sight to a source can scatter low-energy X-rays and mimic the affect of absorption.  The effect of dust scattering is corrected for by multiplying the spectral models by a optically-thick dust scattering term, which has opacity $\tau=0.485\times$\nh (in units of $10^{22}$ cm$^{-2}$) and a scattering halo of 100 times the \emph{Chandra} PSF (see Baganoff et al.\ 2003).} \nh$=7.0_{-0.9}^{+1.1}\times10^{22}$ cm$^{-2}$ for all three sources (the value of \nh is found to vary little between these sources).  About 1\% pileup is expected for the brightest of these sources, which should affect best-fit parameters only negligibly;  all other sources discussed in this study are fainter and will have much smaller amounts of pileup.

The spectrum of the central $15\arcsec$ of the cluster was also extracted and fitted to avoid problems associated with the energy-dependent PSF.  As before, this spectrum can be fitted with an absorbed, one-temperature model with dust-corrected \nh$=8.0^{+1.6}_{-1.2}\times10^{22}$ cm$^{-2}$ and $kT_{\rm{core}}=1.5^{+0.2}_{-0.2}$ keV with $\chi_\nu=0.69$, in the Gehrels statistic. \footnote{http://cxc.harvard.edu/ciao/ahelp/chigehrels.html}  These fits can be improved, particularly in the spectral lines, by the addition of a second component;  this second component can be represented by another thermal model or a powerlaw model with equal success.  A two-temperature thermal model gives a similar fit quality ($\chi_\nu=0.63$) as the one-temperature case.  This fit has a higher absorbing column, \nh$=8.9^{+1.3}_{-1.0}\times10^{22}$ cm$^{-2}$, and two temperatures of $kT_{1}=1.2^{+0.4}_{-0.8}$ keV and $kT_{2}=6.2_{-5.7}$ keV.  For simplicity, the errors are calculated assuming the abundance is equal to the best-fit value of 4.35 times solar, and thus underestimates the true errors.  Alternatively, a one-temperature plus powerlaw model can be fitted to this spectrum well ($\chi_\nu=0.64$).  This fit gives a similar constraint on \nh, a temperature $kT=1.5^{+0.2}_{-0.2}$ keV, and $\Gamma=-2.0_{-2.6}^{+4.5}$ (where flux $\propto \rm{energy}^{-\Gamma}$).  These errors also assume the abundance was frozen at the best-fit value of 5.5 times solar.  Two-temperature models give an absorption-corrected X-ray luminosity, $L_{\rm{X}}(0.5$--$8 \rm{keV})\sim(0.5$--$3)\times10^{35}$ ergs s$^{-1}$ for the cluster, where the range in $L_{\rm{X}}$ is found by refitting the spectra with \nh fixed at the 90\% confidence bounds.  One-temperature/powerlaw models give $L_{\rm{X}}(0.5$--$8 \rm{keV})\sim(0.5$--$1)\times10^{35}$ ergs s$^{-1}$ for the cluster.

The diffuse component, A3, was originally suggested as a candidate ``cluster wind'' (Y02), although it may also be caused by scattered, fluorescent emission.  The X-ray spectrum can be fitted with a power law plus a Gaussian component at 6.4 keV.  This component models Fe K$\alpha$ emission, which is due to radiation scattered from molecular material that fluoresces strongly at 6.4 keV.  While a thermal model can fit the emission, the equivalent widths of the thermal lines are consistent with the ambient GC levels.  This model for the A3 emission could be confirmed with an accurate measurement of the column density of the cloud and its distance from the star cluster (e.g., Sunyaev \& Churazov 1998).  A recent study by Lang, Goss, \& Morris (2002) has claimed that there is molecular material near the Arches cluster, which may be the source of the fluorescence.  A more recent study by Figer et al.\ (2002) measured stellar velocities in the Arches cluster which are substantially different from the that molecular material found nearby.  Alternatively, Cant\'{o} et al.\ (2000) speculated that molecular material from young, low-mass stars in the cluster may be mixed in with the cluster wind; this idea would explain why A3 is centered near the cluster core, instead of being more aligned with the nearby molecular cloud.  If A3 results from scattered emission, it may confuse any search for diffuse emission from the cluster itself.

Two satellite sources, A4 and A5, are symmetrically displaced about the cluster core.  Curiously, the line connecting A4 to A5 is bisected ($69\arcsec$ either side with P.\ A.\ $=85\ddeg6$) by the A1 source in the cluster core.  The orientation of this line is not parallel to any CCD edge, suggesting it is not an instrumental effect.  Although each source only has $\sim35$ background-subtracted counts, their spectra suggest that they are very different sources.  In terms of a color ($C^{4\rm{keV}}=(h-s)/(h+s)$, where $h$ and $s$ are the number of counts from 4--8 keV and below 4 keV, respectively):  $C^{4\rm{keV}}_{A4}=0.92$ and $C^{4\rm{keV}}_{A5}=-0.95$.  The A5 spectrum looks like a common $\sim$1 keV thermal source with an \nh$\sim2.5\times10^{22}$ cm$^{-2}$, while the spectrum of A4 has a highly absorbed \nh$\sim4\times10^{23}$ cm$^{-2}$, and a peak in its spectrum near the iron 6.4 keV line.  The difference in color of the two sources suggests that they have very different extinctions, and are thus not at the same distance; furthermore, the absorbing column towards A5 is significantly less than the canonical GC value.  However, the nature of A4, alone, is peculiar;  could the emission be Fe K$\alpha$ emission?  The A4 source is point-like; this is unusual for scattered flourescent emission in the GC, which tends to be diffuse.  Another possibility could be that the A4 source is a hot ($kT>1$ keV), highly absorbed source with a strong thermal iron line at 6.7 keV.

\subsection{The Quintuplet Cluster}
\label{quint}

\begin{figure}
\plotone{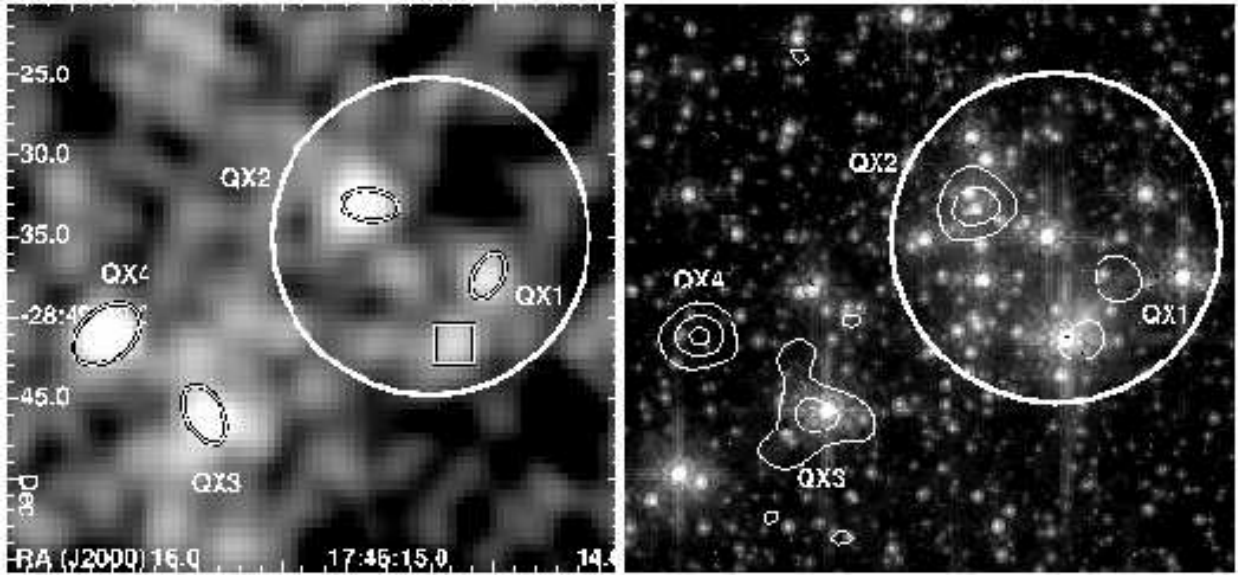}
\caption{(\emph{Left}) \emph{Chandra} X-ray image based on all observations of the Quintuplet cluster, convolved with a $1\arcsec$ Gaussian.  Large circle (in both panels) shows the nominal extent of the cluster core as listed in Table \ref{properties}.  Ellipses show X-ray sources; size of ellipse represents positional uncertainty in the detection.  The square region shows a possible X-ray source that needs to be confirmed; the detection significance of this source is not easily quantified. (\emph{Right}) \emph{HST}/NICMOS image of the same field with the contours from the convolved X-ray image at levels of 8, 15, and 22 counts overlaid. The Pistol star is south of this field of view. \label{quintsmall}}
\end{figure}

The \emph{Chandra} observations presented here are the most sensitive X-ray observations of the Quintuplet to date, and have for the first time detected point sources within the cluster core.  In the central $\sim25\arcsec$ of the cluster, four sources with greater than 2.5 $\sigma$ confidence are detected;  they contribute $\sim$20\% of the flux within the central $50\arcsec\times100\arcsec$.  Figure \ref{quintsmall} shows a comparison of the X-ray emission to near-IR observations with \emph{HST}/NICMOS data (Figer et al.\ 1999b).  The westernmost source, QX1, is the faintest and shows a possible coincidence with source 242 of Figer et al.\ (1999a, hereafter FMM99);  the spectral type of the star is not identified by FMM99.  QX2 is brighter and possibly aligned with source 257 of FMM99, a B0 supergiant.  Next to the east is QX3, within $1\arcsec$ of source 211 of FMM99, a ``Quintuplet proper member'' and tentatively identified as a dusty, late-type WC star.  The easternmost significant X-ray detection near the cluster core (QX4, at $\alpha=$17\shour46\smin14.8\ssec $\delta=$--28\sdeg49\arcmin35\arcsec) does not have a near-IR counterpart.  Moving from east to west the X-ray/IR coincidences become less certain, because of the larger concentration of near-IR sources in the cluster core and typical errors in absolute astrometry of \emph{Chandra} data.  There is potentially a fifth X-ray source, tentatively named CXOGCJ174614.7--284947 and highlighted with a square region in Figure \ref{quintsmall}, within $1\arcsec$ of source 231, another dusty, late-type WC star.  CXOGCJ174614.7--284947 is visible on the merged data set, but not significantly detected on any single observation, making the calculation of a detection confidence nontrivial;  this detection, which needs to be confirmed, has fewer counts than the adjacent QX1 source, giving an upper limit to the detection significance of 2.6 $\sigma$.

Recent high-resolution radio continuum observations with the VLA show several point sources in the Quintuplet cluster (Lang et al.\ 2003).  The emission is generally thermal in nature, originating in the winds of massive stars.  Two of the sources discovered by Lang et al.\ (2003) are aligned with potential X-ray/IR-correlated sources:  QX2 with their QR6 and the tentatively detected CXOGCJ174614.7--284947 with QR7.  The detection of X-ray emission from these stars would be consistent with their identifications made at radio and IR wavelengths.  The Pistol star, one of the most luminous stars known (Figer et al.\ 1999b), is undetected in the X-ray observations to a 3 $\sigma$ detection limit of $\sim$ 1$\times10^{-6}$ counts cm$^{-2}$ s$^{-1}$, giving a conservative constraint on its intrinsic luminosity of L$_{\rm{x}}<10^{33}$ ergs s$^{-1}$.

A spectrum with 112 counts was extracted from the four significantly detected X-ray point sources (see Table \ref{properties}) in the Quintuplet.  Fitting this spectrum with a dust-corrected, absorbed thermal model, with the abundance fixed at twice the solar value, the best-fit values are \nh$=4.3^{+3.3}_{-1.6}$ cm$^{-2}$, $kT=5.1^{+23.5}_{-3.0}$ keV (for an abundance of five times solar, the temperature drops by 0.5 keV).  If the canonical \nh ($6\times 10^{22}$ cm$^{2}$) is assumed, the best-fit gas temperature is $kT=3.3^{+6.3}_{-1.4}$.  After subtracting the contribution from the local background, the flux is found to be $5\times10^{-14}$ ergs cm$^{-2}$ s$^{-1}$.  Correcting for the best-fit extinction and dust-scattering, the X-ray (0.5--8 keV) luminosity is L$_{\rm{x}}=1^{+2}_{-0.2}\times10^{33}$ ergs s$^{-1}$.  Assuming the canonical \nh, the X-ray luminosity is $1.8\times10^{33}$ ergs s$^{-1}$.  The intrinsic luminosity of the Quintuplet is further discussed in \S \ref{discussion}.  To compare the spectra of the individual sources to the total spectrum, the 4 keV color (see \S \ref{arches}) was calculated for each of these sources and the sum of these sources.  For QX1, QX2, QX2, and QX4, the colors are: $C^{4\rm{keV}}=-0.3\pm0.3$,$-0.1\pm0.3$,$0.2\pm0.2$,$0.4\pm0.2$, respectively;  for the total spectrum, $C^{4\rm{keV}}=0.2\pm0.1$.  To first order, the difference in color is caused by different absorption toward each source;  this suggests that the QX1, the only source with a significantly different color, may be a foreground source.  Fortunately, QX1 is the faintest source ($<10$ counts), so it shouldn't significantly bias the best-fit parameters of the total spectrum.

\begin{figure}
\plotone{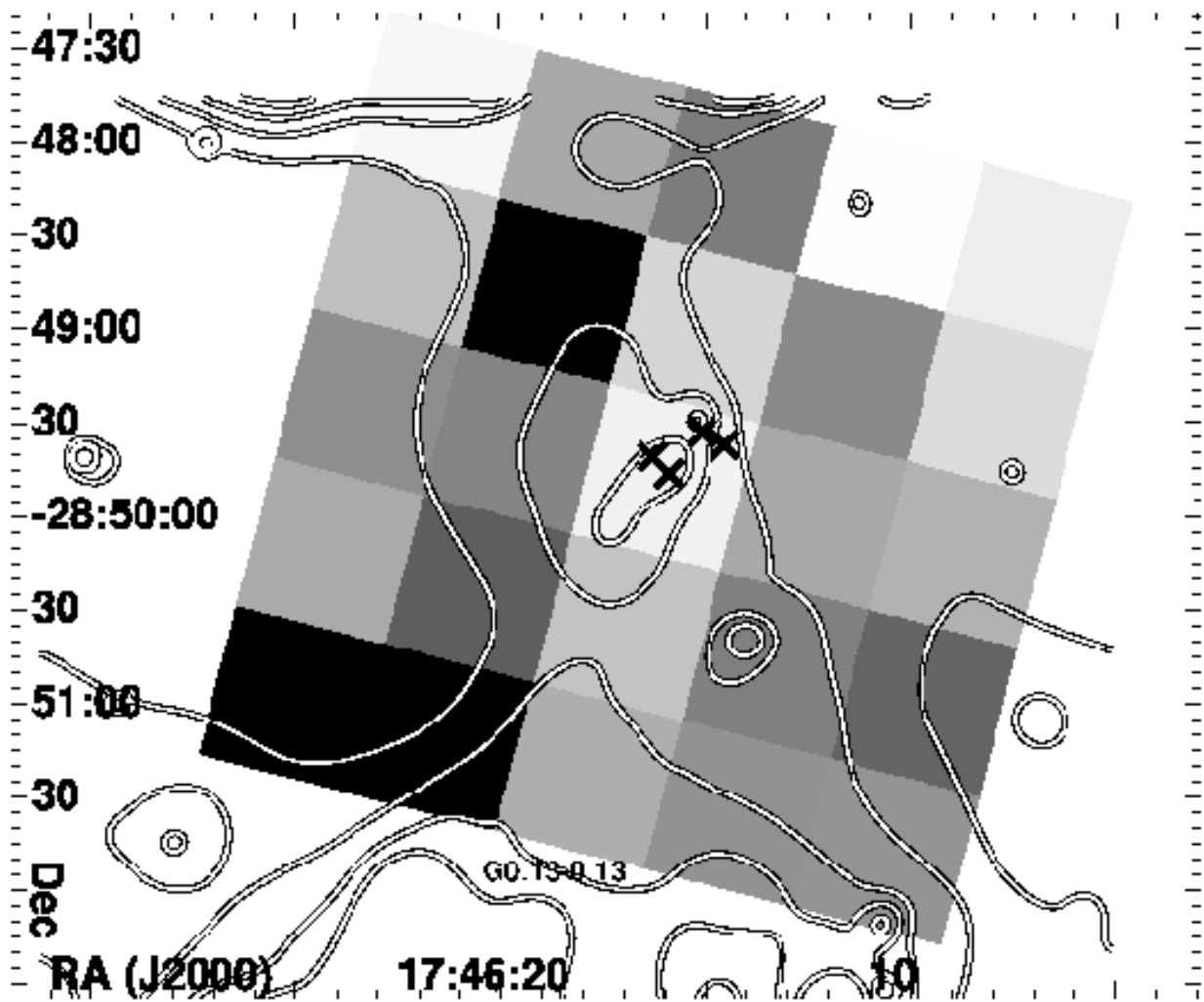}
\caption{Image of gas temperature with contours of X-ray continuum in the vicinity of the Quintuplet cluster.  Grayscale shows the temperature of the best-fit, absorbed thermal model for $50\arcsec$ regions, arranged in a 5 $\times$ 5 matrix.  Spectral response files used for each fit were taken from the center of each region.  Best fit temperature values range from 1.3 (black) to 2.6 (white) keV.  The contours were taken from an adaptively smoothed and exposure-corrected image of the Quintuplet region from 0.5--8 keV.  Point sources with significance $>$2 $\sigma$ were removed prior to smoothing the image and extracting data for the spectral fits.  Crosses show X-ray sources detected near the core of the Quintuplet cluster (QX1 -- QX4); the molecular cloud, G0.13--0.13 (Yusef-Zadeh, Law, \& Wardle 2002), is also labelled to the south.\label{quintbig}}
\end{figure}

Figure \ref{quintbig} shows contours from an adaptively smoothed point-source-free image of the X-ray emission from the Quintuplet cluster overlaid on a map of gas temperature from an array of spectral fits.  The contours show that diffuse emission covers much of the region, with significant peaks at the Quintuplet cluster and toward the Radio Arc/G0.13--0.13 molecular cloud region (Tsuboi, Ukita, \& Handa 1997; Yusef-Zadeh, Law, \& Wardle 2002).  The gas temperature map was made by extracting a series of spectra in a 5 by 5 grid from an event file with point sources of $>$2 $\sigma$ confidence removed;  this analysis is similar to that used to create temperature maps of galaxy clusters (e.g., Kempner, Sarazin, \& Markevitch 2003).  The spectra were fit with an absorbed thermal model plus a Gaussian line at 6.4 keV to account for fluorescent iron emission; metal abundances were frozen at solar values, which is generally consistent with values found when left free to fit.  Spectral response files created for the center of each region were used for the fits, while background events from identical regions of ``blank-sky files'' were modeled and folded into the spectral fits.  Each spectral fit had 600--1000 counts, giving errors on the value of the temperature of about $\pm0.3$ keV.  The values of temperature fit across the Quintuplet region show a peak temperature coincident with the core of the Quintuplet cluster and slighly southeast.  The temperature at this peak is significantly higher than in neighboring regions: $kT=2.42\pm0.5$ keV near the core, compared to an error-weighted average of $kT=1.53\pm0.08$ keV for the neighboring regions.  The absorption-corrected luminosity of the X-ray emission from the central 50\arcsec square region, at the peak in the diffuse emission, is L$_{\rm{x}}=1.9^{+0.6}_{-0.4}\times10^{34}$ ergs s$^{-1}$; the background flux, outside the contours in Figure \ref{quintbig}, is typically half that seen towards the peak in the Quintuplet's diffuse emission, giving a net luminosity, L$_{\rm{x}}\sim1\times10^{34}$ ergs s$^{-1}$. A more precise estimate for the Quintuplet's diffuse flux is limited by confusion with nearby diffuse emission to the north and south.  No 6.4 keV line is significantly detected in the 50\arcsec region coincident with the Quintuplet cluster.

\section{X-ray Emission from Candidate Clusters}
\label{candidate}
Our goal in the following analysis is to use our knowledge of the Arches and Quintuplet to search for new star clusters located near the GC.  The Arches, Quintuplet, and Central clusters are thus generalized as having $L_{\rm{X}}\sim10^{33-35}$ ergs s$^{-1}$, significant smounts (40\%) of diffuse X-ray emission, absorbing columns \nh$\sim4-10\times10^{22}$ cm$^{-2}$, and X-ray sources aligned with windy (e.g., WR and O) stars identified at infrared and radio wavelengths.

\begin{figure}[t]
\plotone{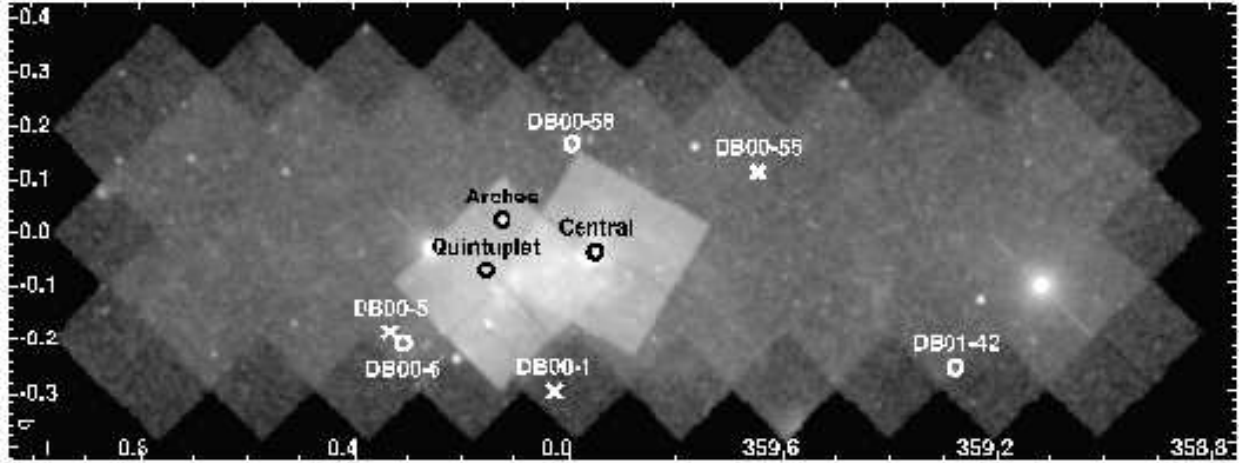}
\caption{X-ray image of 0.5--8 keV emission based on the merged \emph{Chandra} data set used in this study;  observations include those of Sgr A* (Baganoff et al.\ 2003), the Radio Arc (Y02), and the GC survey (Wang, Gotthelf, \& Lang 2002).  Regions show locations of known clusters in black and candidate star clusters in white.  Circles show clusters with X-ray emission;  crosses show those without X-ray emission.  The three major clusters have been observed longer than the candidate clusters.
\label{canddist}}
\end{figure}

Dutra \& Bica (DB00, DB01) searched for massive stellar clusters---using the Arches cluster as a template---by inspecting 2MASS $J$, $H$, and $K_{\rm{s}}$ images.  Follow-up observations with the New Technology Telescope (NTT; Dutra et al.\ 2003) confirmed 42 of their original 57 candidates within 7\sdeg of the GC. Six of these confirmed candidates lie within the central 2\sdeg$\times$0\ddeg8 covered by \emph{Chandra's} Galactic Center Survey (Wang et al.\ 2002).  The distribution of the candidate clusters in the GC is shown in Figure \ref{canddist}, overlaid on an X-ray continnum map of all \emph{Chandra} observations toward the clusters.  In the following sections, we describe the multiwavelength analysis of these candidate clusters.  The clusters are named by the paper in which they are catalogued (DB00 or DB01) and the number they are given in each catalog;  Galactic coordinates are also given for each candidate cluster.  Candidate clusters with associated X-ray emission are described next.

\begin{figure}
\plotone{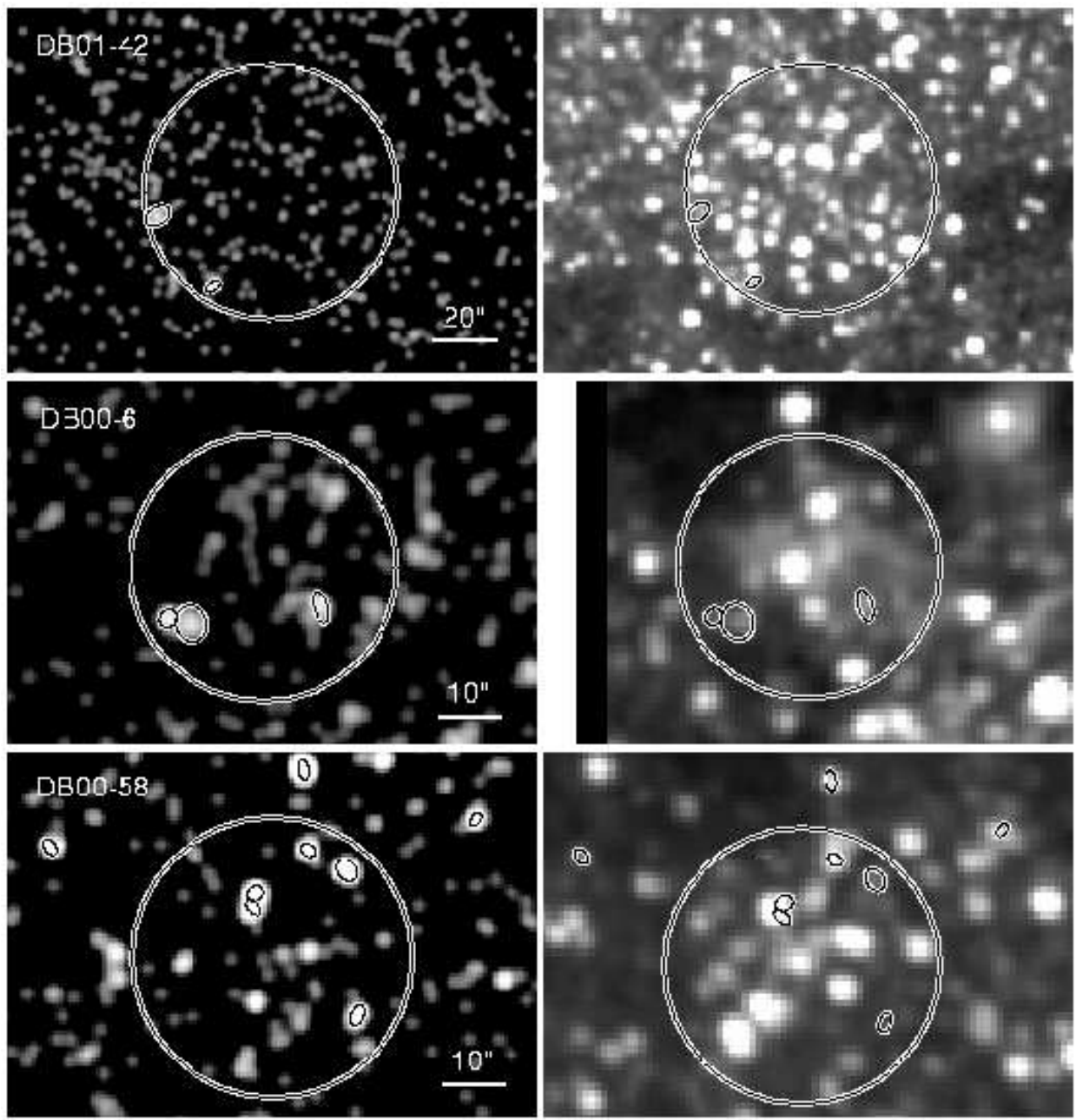}
\caption{\emph{Chandra} and 2MASS $K_s$-band images of the three candidate star clusters with X-ray emission:  DB01-42 (\emph{top panels}), DB00-6 (\emph{middle panels}), and DB00-58 (\emph{bottom panels}).  Regions on the X-ray and IR images are identical for any given cluster candidate.  The large circles shows the size of the cluster core, estimated by eye from the $K_s$ image and as given in Table \ref{properties}; the ellipses show the X-ray sources found within two cluster core radii of the candidate.  All X-ray sources and their potential IR counterparts are listed in Table \ref{xirtab}. \label{candfig}}
\end{figure}

\subsection{DB01-42 (G359.28-0.25)}
\label{db01-42}
\begin{figure}
\plotone{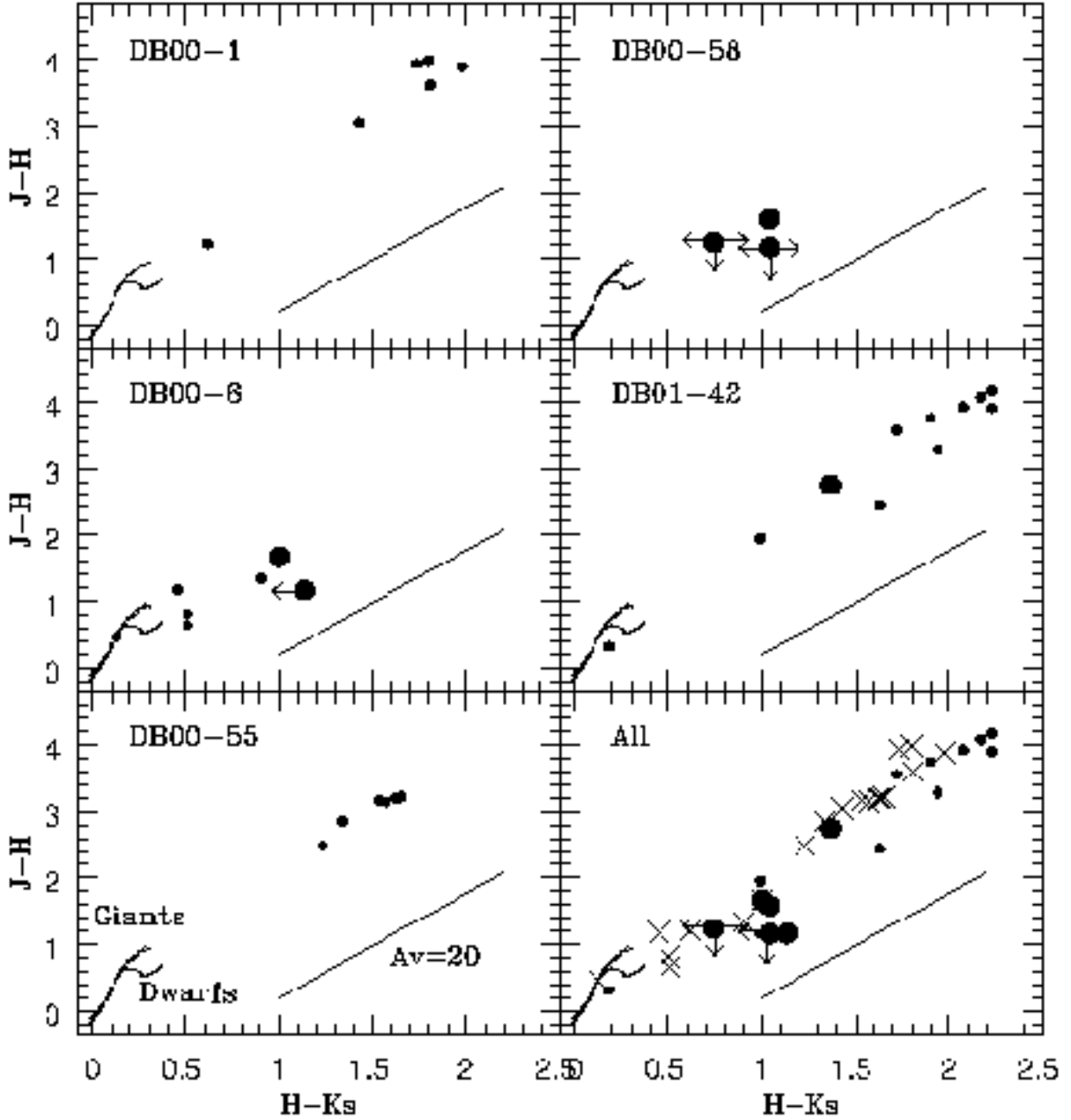}
\caption{\small \emph{(Panels 1--5)} $J-H$ versus $H-K_{\rm{s}}$ for 2MASS sources near candidate clusters.  Large points show IR sources with potential X-ray counterparts, small points show IR sources without an X-ray counterpart.  Arrows indicate upper or lower limits;  all other plotted sources have complete photometry.  Theoretical tracks from Koorneef (1983) are plotted in lower left corners of each panel; a reddening vector is plotted to show the direction that intrinsic colors are shifted by an extinction of \av$=20$ mag, based on Cardelli, Clayton, and Mathis (1989).  Colors converted to 2MASS color system according to Carpenter (2001). Candidate DB00-5 has no 2MASS sources which satisfy our photometry selection criteria (see \S \ref{obs}) and thus is not shown here.  \emph{(Panel 6, bottom right)} $J-H$ versus $H-K_{\rm{s}}$ for \emph{all} 2MASS sources near candidate clusters.  Points show sources from candidate clusters with at least one X-ray/IR correlation with actual X-ray/IR correlations shown with large points, as before.  Crosses show sources from candidates without any X-ray/IR correlations. \label{jhhkfig}}
\end{figure}

As seen in Table \ref{properties}, candidate DB01-42 has the largest angular extent ($40\arcsec$ radius) and is very overdense (1.6 times its background).  Figure \ref{candfig} compares \emph{Chandra} and 2MASS $K_{s}$-band images of the three candidates with X-ray emission.  The near-IR image shows irregular extinction in the south, suggesting that DB01-42 may only appear to be a cluster due to a foreground dust lane.  However, studying the colors of 2MASS sources within the candidate cluster boundary suggests that many of the sources are highly reddened.  Column 6 of Table \ref{properties} shows the number of sources with $H-K_{s}$ colors which are consistent with a GC location;  the vast majority of sources are reddened as if they are near or beyond the GC.  The $J-H$ versus $H-K_{s}$ diagram of Figure \ref{jhhkfig} shows more explicitly how most of the stellar sources with complete photometry require a reddening of \av$>20$ mag to explain their colors.

Of the two X-ray sources within the cluster core, one lies within $1\arcsec$ of a 2MASS source.  The X-ray/IR sources give us two independent methods of measuring the extinction to the cluster, assuming the alignment of the source with the candidate cluster is not coincidental.  X-ray spectral fits are very loosely constrained, but indicate highly absorbed emission.  For a 2 keV thermal plasma model, we find \nh$=(8$--$17)\times10^{22}$ cm$^{-2}$.  This is equivalent to \av$=30$--60 mag, using the conversion factor of \nh/\av$=1.79\times10^{21}$ cm$^{-2}$ mag$^{-1}$ (Predehl \& Schmitt 1995), and considering the effects of optically thick dust scattering.  This range of values for \nh is fairly insensitive to the temperature chosen.  For consistency, this result was compared to the reddening of the near-IR counterpart to the X-ray source, estimated from Figure \ref{jhhkfig}.  By simply measuring the length of the reddening vector required to make the photometry consistent with intrinsic colors, we estimate \av$=20$--30 mag for the X-ray/IR-correlated source.  These values are roughly consistent with each other and suggest that the source is near the GC.

\begin{figure}
\plotone{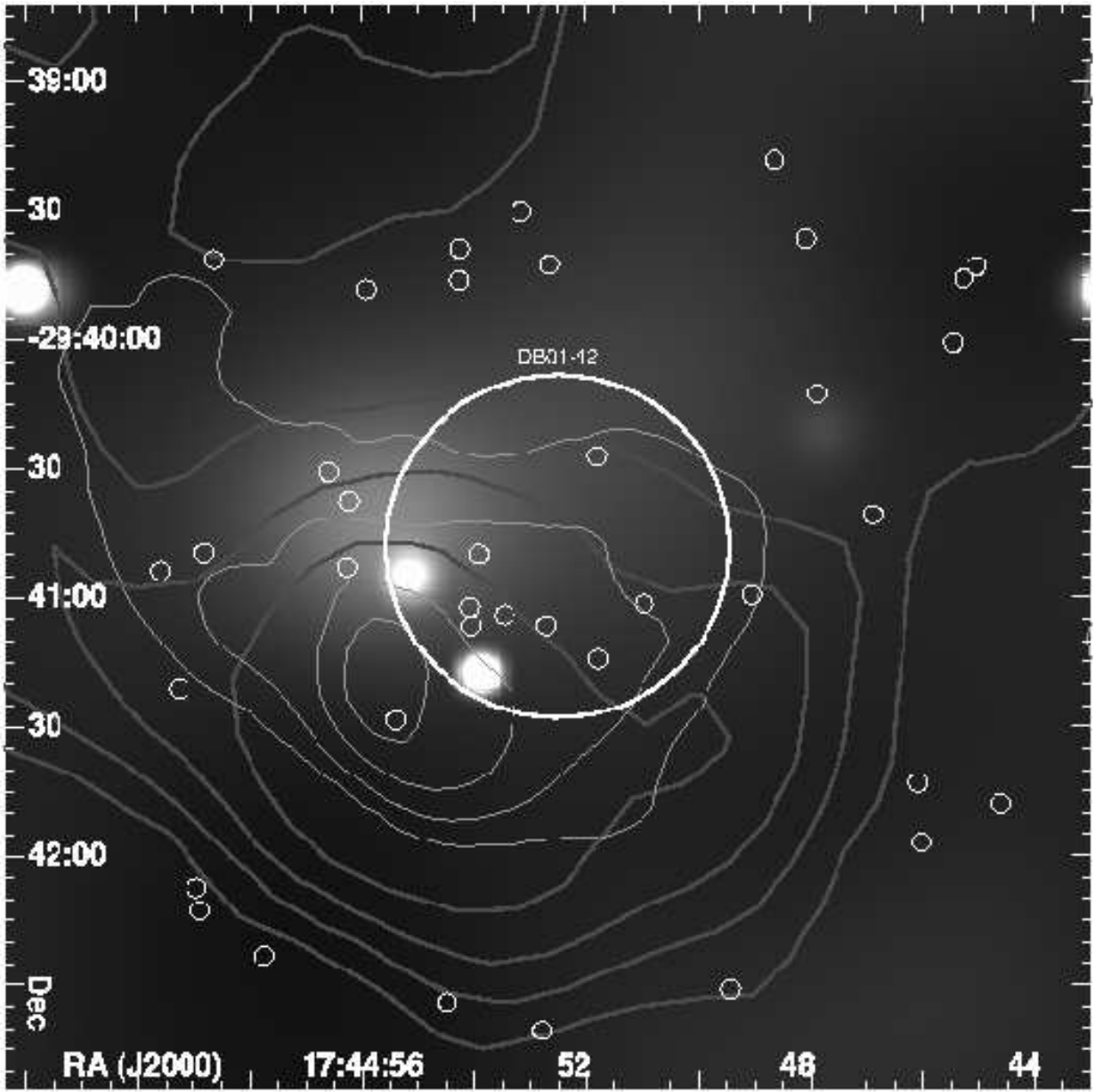}
\caption{Grayscale shows an adaptively-smoothed \emph{Chandra} image of 0.5--8 keV emission from the DB01-42 candidate cluster.  Small circles mark the positions of 2MASS sources with complete photometry and $H-K_{s}>1.4$, which are likely to be in or beyond the GC region (see \S \ref{obs}). The large circle marks the extent of the candidate cluster, DB01-42.  Light contours show 20cm radio emission from the VLA at levels of 0.5, 1, 2, and 4 mJy per 4\arcsec beam; dark contours represent mid-IR emission from the MSX A band at 1.2, 1.6, 2, and $3\times10^{-5}$ W m$^{-2}$ sr$^{-1}$. \label{db01-42fig}}
\end{figure}

The DB01-42 candidate cluster and the X-ray emission may also be associated with the {\small{H II}} region, G359.3--0.3 (Downes et al.\ 1980).  Figure \ref{db01-42fig} shows an adaptively smoothed X-ray image with small circles marking GC-extincted, near-IR sources and contours of mid-IR and 20cm radio emission.  The radio and mid-IR emission have a bow-like shape, with the mid-IR enveloping the radio, which in turn surrounds the candidate cluster and X-ray emission.  The mid-IR is fairly closely aligned with the dust lane which also seems to surround the candidate cluster (see Figure \ref{candfig}).  Downes et al.\ (1980) measured H110$\alpha$ emission and formaldehyde absorption lines towards G359.3--0.3 and found $v_{LSR}^{{\small{H II}}}=-1$ and $v_{LSR}^{H_{2}CO}=-2$ km s$^{-1}$, respectively.  The radial velocity of the {\small{H II}} region is consistent with a foreground location.

\subsection{DB00-5 (G0.33-0.18)}
\label{db00-5}
Candidate DB00-5 appears mostly as diffuse emission in 2MASS images.  Only four point sources were extracted from the 2MASS point source catalog, of which only one had photometry accurate enough to estimate a reddening from the $H-K_s$ color;  this reddening was inconsistent with a GC location.  Recent NTT observation of DB00-5 (see Figure 2 of Dutra et al.\ 2003), shows also a few point sources surrounded by a diffuse feature.  DB00-5 has a mid-IR counterpart, suggesting the presence of thermal emission.  DB00-5 and DB00-6 are located less than $1\arcmin$ from each other, within the Sh2-20 {\small{H II}} region (Sharpless 1959).  If these candidates are located at a similar distance, the low reddening seen toward DB00-5 suggests that DB00-6 may also be located outside the GC.  As discussed below, the reddening estimates toward DB00-6 are similarly small, strengthening the argument for both candidate clusters being foreground objects.

Although seemingly unrelated to the candidate cluster because of its large angular separation, the X-ray source CXOGCJ174708.4--284542, is worth mentioning for its large variability.  The source was observed first on 18 July 2001 at 00:48:28UT;  70 counts were detected in the 10 ks observation, giving a flux of $2.5\times10^{-5}$ counts cm$^{-2}$ s$^{-1}$.  The next observation was the following day 17 July at 14:11:51 UT;  in this observation the source was detected with at only $\sim$2 $\sigma$ significance with about 4 counts, giving a flux of $1.4\times10^{-6}$ counts cm$^{-2}$ s$^{-1}$ (this should be considered an upper limit to the flux, because of the low detection significance).  Thus, the source varied by at least a factor of 15 over 38 hours.  This source is within 0\dasec4 of 2MASS17470848-2845421 with ($J$, $H$, $K_s$)=(13.445, 12.778, 12.580), which places it on the main sequence.  The low reddening toward the source suggests that it is much less than 8 kpc from us.

\subsection{DB00-6 (G0.32-0.21)}
\label{db00-6}
DB00-6 is somewhat larger and denser than the DB00-5 cluster, although it still does not appear to be overdense in the $K_s$ band, as seen in column 5 of Table \ref{properties}.  Column 6 of the same table shows that the reddening of the stars in DB00-6 are mostly inconsistent with GC reddenings.  Similar to DB00-5, DB00-6 has a distinct mid-IR counterpart.

Figure \ref{candfig} shows X-ray and near-IR images of DB00-6 in the middle two panels with X-ray detected sources overlaid on each image.  As shown in Table \ref{xirtab}, two of the three detected X-ray sources have compelling X-ray/IR correlations and are located within the cluster core.  Fitting the X-ray spectrum with an absorbed MEKAL model, we find an X-ray absorbing column of \nh$=(0.6$--$1.6)\times10^{22}$ cm$^{-1}$, which corresponds to \av$\sim5$ mag.  The reddening of the near-IR counterpart (as estimated from the $J-H$ versus $H-K_{s}$ diagram) lies in the range of \av$=$(0--20) mag for the two X-ray/IR sources.  The X-ray and IR values for the extinction toward these sources are inconsistent with a GC location.

\subsection{DB00-58 (G359.99-0.16)}
\label{db00-58}
DB00-58 is a larger, overdense candidate with a mix of GC and non-GC reddened sources.  As seen in Figure \ref{candfig} and Table \ref{xirtab}, this cluster has the most X-ray sources and X-ray/IR correlations of any of the clusters in this study.  Applying the analysis described above, the X-ray absorbing column was measured to be \nh$\sim3\times10^{22}$ cm$^{-2}$, equivalent to \av$\sim15$ mag.  The potential X-ray/IR correlations would require reddenings of \av$=0$--20 mag to make them consistent with unreddened, theoretical values.  Again, these values are inconsistent with the \av of the GC, suggesting that this cluster lies in the foreground of the GC.  Mid-IR emission is detected toward DB00-58 suggesting that it is not an apparent cluster, caused by a region of low extinction.
 
\subsection{DB00-1 (G0.03-0.30)}
\label{db00-1}
The cluster candidate DB00-1 is not detected in the \emph{Chandra} data.  The source is located at the edge of the GC Survey and was only observed for one 10 ks exposure, giving a 3 $\sigma$ limiting flux of $4.9\times10^{-6}$ counts cm$^{-2}$ s$^{-1}$ (see Table \ref{properties}).  Table \ref{properties} shows that this candidate is host to a significant number of near-IR sources with colors consistent with the extinction toward the GC;  this is also seen in the $J-H$ versus $H-K_s$ diagram in Figure \ref{jhhkfig}.  However, DB00-1 is somewhat larger than the known clusters, is less dense, and lacks a mid-IR counterpart.

\subsection{DB00-55 (G359.63-0.09)}
\label{db00-55}
DB00-55, undetected by \emph{Chandra}, is similar to DB00-1 in most respects;  as compared to the known GC clusters, DB00-55 is less dense and lacks a mid-IR counterpart.  However, many of the near-IR sources have colors consistent with GC reddening, as seen in Figure \ref{jhhkfig} or Column (6) of Table \ref{properties}.

\subsection{Probabilities of False Correlation}
\label{falsecorr}
Before conclusions can be reached on the detection of X-ray sources in the candidate clusters, the probability of a false correlation, either having an X-ray source coincidentally fall within a cluster boundary or an X-ray/IR correlation, needs to be addressed.  First, the probability of finding an X-ray source associated with a cluster can be estimated by finding the mean point source density in the \emph{Chandra} GC Survey.  We find 523 sources with $>$ $3\sigma$ significance in the area observed optimally (with two 11 ks observations), which covers a range ($\Delta l$, $\Delta b)=1.75$, $0.4$.  Scaling the count of sources in the survey down to the size of the candidate clusters, the probability of a false correlation is found to be 0.08 for DB00-6, 0.09 for DB00-55, and 0.29 for DB01-42.  However, all three of these clusters have more than one 3 $\sigma$ source within the cluster boundary (with distance $<$r$_{\rm{core}}$).  Since the probability of $n$ coincidental sources is equal to the probability of a single coincidence to the $n$th power, it is unlikely that all X-ray sources found within any single X-ray emitting cluster are coincidental.  A similar analysis is applied to these clusters to estimate the chance of a false correlation between an \emph{Chandra}-detected X-ray source and a 2MASS, near-IR source.  The chance that an X-ray source happens to lie within $1\arcsec$ of a near-IR source in these candidates is 0.19 for DB00-6, 0.25 for DB00-55, and 0.17 for DB01-42.



\section{Discussion}
\label{discussion}

\subsection{X-Ray Sources:  The Arches and Quintuplet Clusters}
The nature of many of the X-ray point sources in the Arches and Quintuplet clusters is made clear by the compelling correlation with near-IR and radio continuum point sources.  The X-ray emission is considered to be produced in shocks formed in the winds of massive stars (O and WR type);  the more luminous of these sources are most likely to be colliding-wind binaries (Stevens \& Hartwell 2003).  Alternatively, for mass-segregated clusters such as the Arches cluster, it is possible that X-ray emission may also arise from accretion onto an intermediate mass black hole, as recent theoretical work suggests (G\"{u}rkan, Freitag, \& Rasio 2003).  

The collision of winds in a binary system can greatly enhance the X-ray luminosity, depending on the type of stars in the binary and the phase of the orbit (Pollock 1987).  It has been suggested that the collective effect of these wind collisions has recently been observed in the Arches cluster, host of the more luminous X-ray sources in the GC region.  Yusef-Zadeh et al.\ (2003) have recently found diffuse nonthermal radio emission from the core of the Arches cluster, coincident with the A1N/S X-ray source.  Although some stellar wind sources can have nonthermal radio emission, this emission is not pointlike;  the source is extended to approximately the size of the cluster core, $9\arcsec$.  This extended nonthermal radio emission is thought to be produced by diffuse shock acceleration (e.g., Bell 1978) in the colliding winds of the cluster.  These energetic electrons may be able to Compton upscatter stellar IR radiation into the X-ray regime.  X-ray spectra from the Arches core (A1N, A1S, and A2 sources) can be adequately fitted with a one-temperature/power-law model;  in this model, the fraction of flux attributed to the power law ($\sim1/6$) is consistent with that expected from inverse Compton scattering.

The scaling of X-ray flux with intrinsic cluster properties can give us another view of the X-ray emission mechanisms.
IR observations have shown that the ratio of mass as well as luminosity between the Arches and Quintuplet is roughly $3\colon1$ (FMM99).  In this study, the ratio of background-subtracted, X-ray flux (0.5--8 keV) from the Arches and Quintuplet is found to be roughly $11\colon1$ ($8\times10^{-5}\colon7\times10^{-6}$ counts cm$^{-2}$ s$^{-2}$);  if only the point source emission is considered, the ratio rises to $18\colon1$.  Correcting this observed flux for absorption is difficult considering the poorly constrained \nh measurement toward the Quintuplet.  From near-IR observations (FMM99; Figer et al.\ 2002), extinctions toward these clusters have been measured as \ak$^{\rm{Arches}}=3.1$ mag and \ak$^{\rm{Quint}}=3.28\pm0.5$ mag (also \ak$^{\rm{QPM}}=2.7$ mag for the ``Quintuplet proper members'').  The near-IR extinctions toward the Arches and Quintuplet are similar within the uncertainties (which are somewhat higher, but still within the errors of the near-IR extinction expected from the best-fit \nh value), so the absorption-corrected X-ray luminosities should scale similarly to the observed X-ray fluxes.  The bulk of the difference in X-ray flux between the Arches and Quituplet is caused by the presence of luminous point sources in the Arches.
Theoretical studies have shown how massive star clusters are subject to processes which are more likely to create X-ray luminous sources.  For example, in N-body simulations, massive stars are found to congregate near the center of their host clusters, where collisions (forming tight binaries or even mergers) are more likely (Portegies Zwart et al.\ 1999).  Massive clusters are much more prone to these ``core collapses,'' making them more likely to be host to massive, X-ray luminous, binary systems and intermediate-mass black holes (G\"{u}rkan, Freitag, \& Rasio 2003).  It may be the proliferation of such systems in the Arches that makes it so much more X-ray luminous than the Quintuplet.

As shown in \S \ref{quint} and Figure \ref{quintbig}, the Quintuplet cluster has diffuse emission and a clear peak in the gas temperature near the cluster core.  In contrast to the Arches diffuse emission, the Quintuplet diffuse emission is clearly thermal, with thermal line emission evident from highly ionized species of S, Si, Ar, Ca, and Fe.  The peak in gas temperature coincident with the cluster core may be due to any or all of the following: (1) an increase in the temperature of truly diffuse emission, (2) an enhancement of nonthermal emission in the cluster core, and (3) the presence of unresolved, discrete sources.  If the emission is truly diffuse, the high temperature in the cluster core suggests that the Quintuplet could be the source of gas heating.  This would be consistent with the model of a cluster wind flow, in which the wind adiabatically cools as it escapes from the Quintuplet cluster (Raga et al.\ 2001).  However, a nonthermal (powerlaw) spectrum has a shape similar a high temperature thermal continuum; the peak in gas temperature may indicate a larger contribution of nonthermal emission in that region.  A third alternative is that the peak in gas temperature at the cluster core could be due to unresolved cluster members.  In this case, hot stellar emission may cause the increase in gas temperature while the outer cooler gas is the genuinely diffuse, ambient gas seen throughout the GC region.  The best-fit temperature in the Quintuplet core is hotter than that expected from single O or WR stars ($<\sim1$ keV; Chlebowski, Harnden, \& Sciortino 1989);  this is also true of the temperature of the emission from the detected X-ray sources in the cluster core.  If this emission is due to unresolved stellar systems, a significant portion of these systems should be colliding wind binaries, which tend to be hotter and/or may emit nonthermal radiation (Pollock 1987).  While young O stars such as $\Theta^{1}$ Ori A can exhibit hard X-ray flares (Schulz et al.\ 2003), such hard X-rays are less common in older, less magnetically active stars such as those found in the Quintuplet cluster (with an age of $3-5$ Myr, as compared to $\sim0.3$ Myr for $\Theta^{1}$ Ori A).  Finally, the luminosity of this nonpoint-source emission from the Quintuplet core is within a factor of two of that of the Arches A3 source (L$_{\rm{x}}^{Q diff}\sim1\times10^{34}$ ergs s$^{-1}$ compared to L$_{\rm{x}}^{A3}\sim1.6\times10^{34}$ ergs s$^{-1}$ according to Y02, or L$_{\rm{x}}^{A3}\sim6\times10^{33}$ ergs s$^{-1}$ if we reanalyze assuming the canonical GC \nh value).

\subsection{X-ray Sources and Extinctions of Candidate Clusters}
\label{canddisc}
Any X-ray source detected in candidate clusters that lie in or beyond the GC would most likely be shocked-wind emission observed from O and WR stars or wind-accreting neutron stars (Muno et al.\ 2003, and references therein).  For example, a limiting flux of $2.2\times10^{-6}$ counts cm$^{-2}$ s$^{-1}$ (typical of our observations of the candidate clusters) could detect a 1 keV thermal source in the GC with a luminosity, L$_{\rm{x}}\sim5\times10^{33}$ ergs s$^{-1}$;  this temperature and luminosity is similar to that of colliding wind binary systems (e.g., Portegies Zwart, Pooley, \& Lewin 2002).  Nonthermal sources tend to be spectrally hard, and thus are more likely to be detected in the GC.  A $\Gamma=2$ powerlaw would be detectable at our typical flux limit, if it had a luminosity, L$_{\rm{x}}\sim5\times10^{32}$ ergs s$^{-1}$.  Candidate DB01-42 has an \nh that is consistent with a GC location, so its X-ray sources have an origin in one of the above mechanisms.  Of the two X-ray sources in DB01-42,  one has a counterpart with $K_{s}\sim13$ mag and another has no counterpart detected by 2MASS, down to the local detection limit of $K_{s}\sim13$; this is consistent with Pfahl, Rappaport, \& Podsiadlowski (2002), which finds that most neutron star companions should have intrinsic magnitudes of $K>11$ mag, or $K>13$ mag for a typical GC-like extinction.  The flux limit of this survey is adequate to detect the more luminous colliding-wind binary systems in the GC (Pollock 1987).  These systems tend to consist of early-O or WR type stars, which would have $K$-band magnitudes brighter than 13th magnitude, with the GC extinction and distance.  Spectrally soft sources in the GC would need to be dramatically more luminous (L$_{\rm{x}}>10^{34}$ ergs s$^{-1}$) to be detected in this survey; sources at our detection limit which are nearer than the GC will be much less luminous and are more likely to be CVs, main sequence stars, or young stellar objects.

Measuring the extinction toward these clusters can help us determine where they are with respect to the GC.  As shown in \S \ref{candidate}, candidate DB01-42 has associated X-ray emission and possible X-ray/IR correlated sources, which allows a comparison of two independent measurements of the cluster's extinction.  The X-ray extinction for this candidate is found to be consistent with typical GC values;  the IR reddening of the X-ray/IR source is somewhat less, but consistent within the errors to a GC reddening.  Figure \ref{jhhkfig} and Table \ref{properties} reinforce this point by showing that the majority of the IR sources near this cluster candidate are also reddened like the GC.

DB00-1 and DB00-55 have no X-ray emission with which to measure an X-ray extinction.  Interestingly, the reddening estimated from near-IR point sources show that the candidates are extincted as if they were in the GC.  Color-color diagrams from regions near these candidates do not show the same density of highly extincted sources, validating the idea that these candidates are distinct associations and are located in the GC.

Two of the candidates with X-ray emission (DB00-6 and DB00-58) have X-ray absorbing columns inconsistent with typical GC values.  A third candidate, undetected in X-rays (DB00-5), is located in the same {\small{H II}} region as a low-absorption candidate (DB00-6), suggesting that it too is not in the GC.  Figure \ref{jhhkfig} shows that the X-ray/IR sources for these clusters have near-IR extinctions that are relatively small (\av$<20$).  We can also see this by comparing the number of sources with and without GC-like reddening (Col. [6] of Table \ref{properties}); Candidates DB00-5, DB00-6, and DB00-58 all have the highest fraction of sources with non-GC-like $H-K_s$ colors.


%

\subsection{New GC clusters?}
\label{newgcc}
The known GC clusters are similar in four of the diagnostics presented in Table \ref{properties}:  density, size, reddening, and mid-IR emission.  No candidate cluster has all these characteristics to the degree seen in the known clusters.  Using the simplest of these tests, extinction, three candidate clusters were found unlikely to be in the GC, either directly (DB00-6, DB00-58), or by association with other non-GC candiates (DB00-5).

Only one of the candidates with X-ray emission is likely located in the GC:  DB01-42.  With an observed flux of $\sim5\times10^{-6}$ counts cm$^{-2}$ s$^{-1}$, highly extincted X-ray and near-IR emission, this candidate resembles the Quintuplet cluster (although they are much different in angular size: $12\arcsec$ as compared to $40\arcsec$).  The coincidence of DB01-42 inside the bow-shaped {\small{H II}} region, G359.3--0.3, suggests that the candidate cluster may be the source of the ionization and dust heating seen in radio and mid-IR images.  Although the X-ray and near-IR emission clearly originate in the GC, the velocity of the {\small{H II}} region is unusually small for the GC, suggesting that G359.3--0.3 may not be in the GC.  Regardless of the possible correlation with G359.3--0.3, X-ray and near-IR emission indicate that DB01-42 is likely to be a new GC cluster.  Future X-ray observations could search for diffuse emission, a tell-tale signature of young and/or massive stars.

Finally, candidates DB00-1 and DB00-55 have no X-ray emission, but their near-IR sources have GC-like reddening;  this suggests that these clusters are likely to be near the GC, but X-ray data are not sensitive enough to detect them.  The Arches and Quintuplet clusters have background-subtracted fluxes, $F_{X}(0.5$--$8$keV$)=9\times10^{-5}$ and $7\times10^{-6}$ counts cm$^{-2}$ s$^{-1}$, respectively;  these fluxes are larger than the X-ray flux limits of both undetected candidate clusters (see Table \ref{properties}).  If the candidate clusters were as luminous and extincted as the Arches and Quintuplet, then the candidates should have been detected by these observations.  However, it may be possible that the candidates have X-ray luminosities like the Quintuplet (the fainter of the known clusters), but are more extincted.  Simple spectral simulations show that if candidate DB00-1 had a luminosity like the Quintuplet, but an \nh $>1.5$ times greater, then the cluster would not be detected by current observations.  By the same analysis, DB00-55 could have the Quintuplet's flux and not be detected if \nh was $>$3 times higher than the Quintuplet's.  It is more likely that the undetected candidate clusters are intrinsically less luminous than the Quintuplet, and thus they are probably less massive or are not host to any of the massive stars found in the Quintuplet.  These candidates are also the only two clusters not found to have any mid-IR emission, suggesting that they are somewhat older than the other clusters in this study; in fact, Dutra et al.\ (2003) suggested this for candidate DB00-1, but not DB00-55.  One possibility is that DB00-1 and DB00-55 are stellar clusters in the process of being dissolved by the GC's strong tidal gravitational field.  However, all we can conclusively say is that these clusters seem to be significantly less luminous than the known, massive, GC clusters.

Portegies Zwart et al.\ (2002) predicted 10--50 clusters similar to the known, massive clusters in the central 200 pc ($\pm1$\sdeg), at varying stages of dissolving into the stellar background.  The near-IR searches of DB00 and DB01 found many candidates and Dutra et al.\ (2003) confirmed that most of them are stellar clusters, if not necessarily in the GC.  Of the six of these candidates which lie in the central 2\sdeg$\times0\ddeg8$, we have found that:  three are \emph{not} located in the GC, two \emph{are} in the GC but are much less luminous than known GC clusters, and one is a reasonable candidate for a Quintuplet-like GC cluster or association.  Further X-ray and spectroscopic near-IR observations are necessary to solidify the association of these candidates with the GC and to search for the unusual stellar signatures seen in known GC clusters.  We also note that the simple model of Portegies Zwart et al.\ (2002) predicts several more clusters within the central 200 pc;  further searches of 2MASS and DENIS data archives may prove fruitful.

\acknowledgments
This work was made possible by grant NASA8-39073.  This research has made use of the NASA/IPAC Infrared Science Archive, which is operated by JPL/CalTech, under contract with NASA, and NASA's Astrophysics Data System.  We thank Fred Baganoff, Mark Wardle, Jessica Law, and the \emph{Chandra} X-ray Center for assistance and helpful discussions throughout this work.  We also thank the anonymous referee for their valuable comments.


\begin{deluxetable}{lcccccccc}
\tabletypesize{\scriptsize}
\tablecaption{Properties of Known and Candidate Clusters \label{properties}}
\tablehead{
\colhead{} & \colhead{R.A.} & \colhead{Dec.} & \colhead{r$_{core}$\tablenotemark{a}} & \colhead{} & \colhead{non-GC color/GC color\tablenotemark{c}} & \colhead{} & \colhead{X-flux limit\tablenotemark{e}} & \colhead{} \\
\colhead{Candidate} & \colhead{(J2000)} & \colhead{(J2000)} & \colhead{(arcsec)} & \colhead{$\rho_{core}$/$\rho_{bg}$\tablenotemark{b}} & \colhead{(number of srcs/number of srcs)} & \colhead{ObsIDs\tablenotemark{d}} & \colhead{(counts cm$^{-2}$ s$^{-1}$)} & \colhead{\emph{MSX}} \\
\colhead{1} & \colhead{2} & \colhead{3} & \colhead{4} & \colhead{5} & \colhead{6} & \colhead{7} & \colhead{8} & \colhead{9}
}
\startdata
Arches  & 17\shour45\smin50\dsec4 & --28\sdeg49\arcmin20\arcsec & 10 & 1.6 & 1/1  & 945,2276,2284 & 1.1e--6 & yes \\
Quintuplet&17\shour46\smin14\dsec8& --28\sdeg49\arcmin35\arcsec & 12 & 1.6 & 0/8  & 945,2273,2276 & 5.4e--7 & yes \\
Central   & 17\shour45\smin40\dsec1 & --29\sdeg00\arcmin28\arcsec & 10 & 1.6 & 0/3  & 242 & 7.3e--7 & yes \\
\tableline
DB00-1  & 17\shour46\smin51\dsec2 & --29\sdeg3\arcmin47\arcsec & 18 & 1.2 & 1/9 & 2282 & 4.9e--6 & no \\
DB00-5  & 17\shour47\smin7\dsec0  & --28\sdeg46\arcmin4\arcsec & 10 & 0.4 & 1/0 & 2273,2288 & 2.2e--6 & yes \\
DB00-6  & 17\shour47\smin9\dsec4  & --28\sdeg46\arcmin26\arcsec & 21 & 0.8 & 9/1 & 2273,2288 & 2.2e--6 & yes \\
DB00-55 & 17\shour44\smin24\dsec7 & --29\sdeg12\arcmin15\arcsec & 17 & 1.0 & 0/7 & 2267,2268 & 2.1e--6 & no \\
DB00-58 & 17\shour44\smin59\dsec9 & --28\sdeg51\arcmin37\arcsec & 22 & 1.7 & 3/7 & 2284,2287 & 2.1e--6 & yes \\
DB01-42 & 17\shour44\smin52\dsec5 & --29\sdeg40\arcmin48\arcsec & 40 & 1.6 & 2/40 & 2270,2278 & 2.2e--6 & yes \\ 
\enddata
\tablenotetext{a}{Radius of core of cluster candidate, estimated by eye from 2MASS K$_{s}$-band images.}
\tablenotetext{b}{Contrast of cluster core, as measured by the relative number of K$_{s}$-band sources in the core relative to the number in an annulus from 1--3 times the core radius.}
\tablenotetext{c}{The numbers of sources with $H-K_{s}$ colors inconsistent/consistent with GC values.  The range of colors chosen correspond to the reddened colors of the bluest and reddest sources expected to be visible at the typical extinction toward the central degree, \ak$=2$.  See \S \ref {obs} for details.}
\tablenotetext{d}{\emph{Chandra} ObsIDs which observed the given cluster.}
\tablenotetext{e}{X-ray flux measured across the entire calibrated \emph{Chandra} band, 0.3--10 keV.}
\end{deluxetable}

\clearpage

\begin{deluxetable}{cccccc}
\tabletypesize{\scriptsize}
\tablecaption{\emph{Chandra} Point Sources Detected Toward Known and Candidate Clusters \label{xirtab}}
\tablehead{
\colhead{} & \colhead{} & \colhead{X-ray flux (sig)} &  \colhead{distance\tablenotemark{b}} & \colhead{} & \colhead{X-ray/IR offset} \\
\colhead{Cluster} & \colhead{X-ray source\tablenotemark{a}} & \colhead{(counts cm$^{-2}$ s$^{-1}$)} &  \colhead{(arcsec)} & \colhead{IR counterpart\tablenotemark{c}} & \colhead{(arcsec)} \\
\colhead{(1)} & \colhead{(2)} & \colhead{(3)} &  \colhead{(4)} & \colhead{(5)} & \colhead{(6)}
}
\startdata
Arches   & {\bf{174550.3--284922}} (A1S) & 3.4e-5 (30.0 $\sigma$) & 2 & F02--6 & 0.8 \\
--       & {\bf{174550.5--284919}} (A1N) & 3.0e-5 (27.9 $\sigma$) & 3 & F02--7 & 0.5 \\
--       & {\bf{174550.2--284911}} (A2) & 2.8e-5 (27.3 $\sigma$) & 8 & F02--9 & 0.6 \\
--       & 174549.4--284918 & 3.5e-6 (3.9 $\sigma$) & 13 & $\ldots$ & $\ldots$ \\
Quintuplet&174614.5--284937 (QX1) & 1.3e-6 (2.6 $\sigma$) & 5  & FMM99--242 & 0.4 \\
--       & {\bf{174615.0--284933}} (QX2) & 1e-6 (3.7 $\sigma$) & 5  & FMM99--257 & 0.6 \\
--       & 174615.8--284946 (QX3) & 1.4e-6 (6.2 $\sigma$) & 18 & FMM99--211 & 0.7 \\
--       & 174616.3--284941 (QX4) & 2e-6 (7.6 $\sigma$) & 21 & $\ldots$ & $\ldots$ \\
Central\tablenotemark{d} & {\bf{174539.7--290029}} & 2.1e-7 (5.1 $\sigma$) & 3 & IRS13E & 0.6 \\
\tableline
DB00-1   & $\ldots$ & $\ldots$ & $\ldots$ & $\ldots$ & $\ldots$ \\
DB00-5   & 174708.4--284542 & 2.5e-5 (13.2 $\sigma$) & 29 & 17470848--2845421 & 0.4 \\
DB00-6   & 174708.7--284632 & 2.2e-6 (2.9 $\sigma$) & 11 & 17470872--2846317 & 0.4 \\
--       & {\bf{174710.2--284634}} & 2.9e-6 (4.0 $\sigma$) & 14 & 17471020--2846346 & 0.7 \\
--       & {\bf{174710.5--284633}} & 2.2e-6 (3.4 $\sigma$) & 16 & 17471050--2846339 & 0.2 \\
DB00-55  & $\ldots$ & $\ldots$ & $\ldots$ & $\ldots$ & $\ldots$ \\
DB00-58  & 174500.1--285127 & 4.3e-6 (6.6 $\sigma$) & 10 & 17450021--2851282 & 1.9 \\
--       & 174500.1--285129 & 2.7e-6 (4.3 $\sigma$) & 10 & 17450021--2851282 & 1.5 \\
--       & {\bf{174458.9--285145}} & 2.1e-6 (3.3 $\sigma$) & 15 & 17445890--2851460 & 0.2 \\
--       & {\bf{174459.4--285120}} & 2.1e-6 (3.3 $\sigma$) & 17 & 17445951--2851205 & 0.3 \\
--       & {\bf{174459.0--285123}} & 4.2e-6 (6.5 $\sigma$) & 17 & 17445905--2851236 & 0.4 \\
--       & 174459.5--285108 & 6.1e-6 (9.6 $\sigma$) & 29 & 17445953--2851079 & 0.1 \\
--       & 174457.5--285115 & 1.8e-6 (2.9 $\sigma$) & 37 & 17445755--2851160 & 0.7 \\
--       & 174502.4--285120 & 1.8e-6 (2.9 $\sigma$) & 38 & 17450254--2851200 & 0.2 \\
DB01-42  & {\bf{174455.1--294055}} & 2.8e-6 (3.5 $\sigma$) & 36 & 17445511--2940553 & 1.0 \\
--       & 174453.8--294117 & 2.1e-6 (3.2 $\sigma$) & 36 & 17445405--2941179 & 2.4 \\
\enddata
\tablenotetext{a}{Official source names are preceeded by ``CXOGCJ'';  common names are shown in parentheses.  Sources shown in {\bf Boldface} have the following properties: (1) 3 $\sigma$ confidence in detection, (2) located within estimated cluster core, as listed in Table \ref{properties}, and (3) has IR counterpart within $1\arcsec$ of X-ray source position.}
\tablenotetext{b}{Angular distance of X-ray source from center of cluster.}
\tablenotetext{c}{IR counterparts found from the literature and archives.  F02, Figer et al.\ (2002); FMM99, Figer et al.\ (1999b); IRS13E is discussed by Coker, Pittard, \& Kastner (2002).  All other sources from the 2MASS point source catalog; official names are preceeded by ``2MASS''.}
\tablenotemark{d}{Baganoff et al.\ (2003) also describes potential correlations of X-ray sources with stars AF NW (Paumard et al.\ 2001) and IRS 16SW.  The most complete photometry of the region is given by Muno et al.\ (2003), which shows four point sources in the central 10\arcsec.}
\end{deluxetable}

%
%

\end{document}